\documentclass[a4paper,11pt]{article}
\usepackage{pos}
\usepackage{subfigure}
\newcommand*\diff{\mathop{}\!\mathrm{d}}
\newcommand{\Kceil}{\lceil K \rceil }
 \newcommand{\GeV}{{{\,}\textrm{GeV}}}

\usepackage{braket}
\usepackage{bm}

\title{Quark jet evolution: from classical to quantum simulation}

\author*[a]{Meijian Li}

\affiliation[a]{Instituto Galego de Fisica de Altas Enerxias (IGFAE), Universidade de Santiago de Compostela, E-15782 Galicia, Spain}

\emailAdd{meijian.li@usc.es}

\abstract{
Quark jet provides one of the best ways to probe the matter produced in ultrarelativistic high-energy collisions, from cold nuclear matter to the hot quark-gluon plasma. 
In this proceeding paper, we review a series of works on the development of nonperturbative computational framework of in-medium quark jet evolution, from classical to quantum simulation. 
The application of the time-dependent Basis Light-front Quantization (tBLFQ), a nonperturbative computational approach based on light-front Hamiltonian formalism, to in-medium jet evolution enables a fully quantum treatment to the jet state on the amplitude level. 
Based on the tBLFQ framework, with applying novel quantum technologies, we have constructed a digital quantum circuit that tracks the evolution of a multi-particle jet probe within a stochastic color background field.
With the obtained simulation results, we extracted the medium induced modification in terms of jet momentum broadening and gluon production.
These studies provide a baseline for future works of in-medium jet evolution using quantum computers.
}

\FullConference{%
  12th Edition of the Large Hadron Collider Physics conference (LHCP 2024)\\
  3-7 June 2024 \\
  Boston, USA 
}

\begin{document}
\maketitle
\section{Introduction}
In high-energy collisions, a jet is a collimated beam of particles produced by the splitting of a common ancestor (quark or gluon).
The jet traverses through and interacts with the surrounding medium matter that is also produced in the collisions, making it an ideal probe of matter and an effective tool to understand interactions. 
In our theoretical study and simulations~\cite{Li:2020uhl, Li:2021zaw, Li:2023jeh, Barata:2022wim, Barata:2023clv}, we focus on the relatively early stages of the quark jet where it can be treated as an energetic QCD states consist of single quark and quark-gluon components.
We study the evolution of the jet state in the presence of a background gluon field, a fundamental process that is applicable to study various different scenarios, ranging from deep inelastic scattering, proton-nucleus scattering, to heavy ion collisions.
In the following, we first introduce the classical simulation framework in Sec.~\ref{sec:classical}, then the quantum simulation algorithm developed from it in Sec.~\ref{sec:quantum}. We present and discuss selected simulation results in Sec.~\ref{sec:res}, and we summarize in Sec.~\ref{sec:sum}.

\section{Jet evolution in time-dependent Basis Light-Front Quantization (tBLFQ)}\label{sec:classical}
The time-dependent Basis Light-Front Quantization (tBLFQ)~\cite{1stBLFQ} is a nonperturbative computational approach to solve time-dependent  problems in light-front Hamiltonian formalism with the implementation of a basis function representation.
In studying jet evolution, we developed the first tBLFQ computational framework for a quark jet as a $\ket{q}$ state traversing a classical color background field in Ref.~\cite{Li:2020uhl}, then we extended the framework to the $\ket{q} + \ket{qg}$ Fock space~\cite{Li:2021zaw, Li:2023jeh}. 
The computation is carried out on classical computers. The employed theoretical method performs real-time evolution of the quantum state at the amplitude level, making it well-suited for adaptation to quantum simulation.

We consider a high-energy quark jet moving in the positive $z$ direction, traversing a medium moving in the negative $z$ direction. 
We treat the quark as a quantum state and the medium as an external background field $ \mathcal A^\mu_a$, with the interaction occurring over a finite distance of $0\le x^+\le L_\eta$. 
The light-front Hamiltonian can be derived from the QCD Lagrangian with the replacement $A^\mu_a\to A^\mu_a+ \mathcal A^\mu_a$, with $A^\mu $ denoting the gauge field. 
In the truncated Fock space $\ket{q}+\ket{qg}$, the resulting Hamiltonian consists of three parts, the kinetic energy, the interaction between the quark and the dynamical gluon, and the medium interaction,
  \begin{align}\label{eq:H}
      P^-(x^+)=P_{KE}^- + V_{qg}+V_{\mathcal{A}}(x^+)
         \;.    
   \end{align}
We adopt the McLerran-Venugopalan (MV) model~\cite{McLerran:1993ni} in formulating the background field $\mathcal{A}^\mu$. It is a classical field satisfying the reduced Yang-Mills equation,
\begin{align}\label{eq:poisson}
 (m_g^2-\nabla^2_\perp )  \mathcal{A}^-_a(\vec{x}_\perp,x^+)=\rho_a(\vec{x}_\perp,x^+)\,,\quad
 \braket{\rho_a(x)\rho_b(y)}=g^2\tilde\mu^2\delta_{ab}\delta^2(\vec{x}_\perp-\vec{y}_\perp)\delta(x^+-y^+)\;,
\end{align}
in which $m_g$ is the infrared regulator. 
The saturation scale is $Q_s^2=C_F (g^2\tilde\mu)^2L_\eta/(2\pi)$, with $C_F=(N_c^2-1)/(2N_c)$.

 The fermion and gauge fields are quantized on a discrete 3-dimensional momentum basis space. Correspondingly, each single-particle mode contains five quantum numbers, 
\begin{align}\label{eq:basis}
 \beta_l = \{k^+_l, k^x_l, k^y_l, c_l, \lambda_l \}, \text{ with } l=q, g \;,
\end{align}
in which $\{k^+, k^x, k^y\}$ are the dimensionless three-momentum quanta, $\lambda$ the light-front helicity, and $c=1,2,\ldots, N_c$ ($a=1,2,\ldots N_c^2-1 $) the color index of the quark (gluon). 
The periodic transverse lattice extends from $-L_\perp$ to $L_\perp$ over $2N_\perp$ sites for each side. 
The corresponding momentum space is also a periodic lattice of  $2N_\perp$ sites with spacing $d_p\equiv \pi/L_\perp$, such that $(p^x, p^y) = (k^x, k^y)d_p $.  
The longitudinal momentum $p^+$ is quantized in units of $2\pi/L$, and the quanta $k^+$ takes a positive (half-)integer number for gluon(quark) . The total longitudinal momentum quanta of each Fock state is given by $K$, a half-integer. 
The multi-particle state can be constructed as a tensor product of single-particle modes, e.g., $\ket{\beta_{qg}}=\ket{\beta_q}\otimes\ket{\beta_g}$. The total number of basis states for the $\ket{q} +\ket{qg}$ space  is therefore $N_{\text{tot}} = 2 N_c (2 N_\perp)^2 + 4 N_c(N_c^2-1) \lfloor K\rfloor (2 N_\perp)^4$.
In such discrete-momentum basis representation, the jet state can be written as a superposition of different basis states,
\begin{align}\label{eq:basis_expansion}
  \ket{\psi;x^+}=\sum_\beta v_{\beta} (x^+) \ket{\beta}\;.
\end{align}
 The column vector of basis coefficients $v_{\beta} (x^+) $ is the light-front wavefunction of the jet. 
 
The jet state obeys the evolution equation on the light front, 
\begin{align}
  \label{eq:ShrodingerEq}
  i\frac{\partial}{\partial x^+}\ket{\psi;x^+}=\frac{1}{2}P^-(x^+)\ket{\psi;x^+}\;. 
\end{align}
We decompose the time-evolution operator into many small steps of the light-front time $x^+$, and let the operation of each timestep act sequentially to the initial wavefunction,
\begin{align}\label{eq:time_evolution_exp}
 \begin{split}
  \ket{\psi;x^+}=\mathcal{T}_+ &e^{-\frac{i}{2}\int_0^{x^+}\diff z^+P^-(z^+)}\ket{\psi;0}
   =\lim_{n\to\infty}\prod^n_{k=1}
   \underbrace{e^{-\frac{i}{2}\int_{x_{k-1}^+}^{x_k^+}\diff z^+P^-(z^+)}}_{U_k}
   \ket{\psi;0}
 \;,
 \end{split}
\end{align}
with $\mathcal{T}_+$ denoting light-front time ordering. The step size is $\delta x^+ \equiv x^+/n$, and the intermediate time is $x_k^+=k\delta x^+ (k=0,1,2,\ldots,n)$ with $x_0^+=0$ and $x_n^+=x^+$.
Within each short time step, the Hamiltonian can be considered time-independent, then the evolution operator can be approximated using a product formula.
An efficient numerical method based on the matrix structures of the operators is given in Ref.~\cite{Li:2021zaw}, written schematically,
   \begin{align}\label{eq:product_formula}
 \begin{split}
   \lim_{\delta x^+\to 0}U_k
  & =e^{-\frac{i}{2} P^-(x_k^+) \delta x^+}
  \approx 
    [\mathcal{FT}] e^{-\frac{i}{2} V_{\mathcal{A}}(x_k^+)\delta x^+}[\mathcal{FT}^{-1}]
      e^{-\frac{i}{2} (V_{qg}+P^-_{\text{KE}})\delta x^+} 
 \;.
 \end{split}
\end{align}  
Here, $ \mathcal{FT}$ ($ \mathcal{FT}^{-1}$ ) is the (inverse-)Fourier Transform between the transverse position and momentum space.
Note that the computation in Ref.~\cite{Li:2021zaw} is carried out in the interaction picture, such that the kinetic energy term enters as phase factors. The operation of $V_{\mathcal A}$ is carried out by matrix exponentiation in the coordinate space, and that with $V_{qg}$ uses the fourth-order Runge-Kutta method in the momentum space. 
By obtaining the light-front wavefunction of the evolved state, it is straightforward to measure observables of interest, $\braket{\psi;x^+ |\hat O|\psi; x^+ }$.

\section{Quantum simulation algorithm of jet evolution}\label{sec:quantum}

Based on the tBLFQ computational framework of quark-nucleus scattering in Refs.~\cite{Li:2020uhl, Li:2021zaw}, and the quantum strategy proposed for computing jet quenching parameter in Ref.~\cite{Barata:2021yri}, we developed a quantum simulation algorithm jet evolution. First for the single quark state (the $\ket{q}$ Fock sector) in Ref.~\cite{Barata:2022wim}, then we extended the algorithm to the $\ket{q}+\ket{qg}$ Fock space~\cite{Barata:2023clv}.
The developed digital quantum simulation algorithm can be summarized by following the five generic steps:
\begin{enumerate}
    \item Define the problem Hamiltonian. The Hamiltonian of jet evolution is given in Eq.~\eqref{eq:H}, and its matrix element form in the chosen basis space can be found in ~\cite{Li:2021zaw}. 
    Here, we make two simplifications to reduce the usage of computational resources, $N_c=2$, and all particles having spin up configurations.
    \item Basis encoding. 
    We map the basis states as written in Eq.~\eqref{eq:basis} to the qubit state according to the binary representation of the quantum numbers.
    We encode an arbitrary basis state in the $\ket{q}$ or $\ket{qg}$ sector in the following way,
\begin{align}\label{eq:encoding_complete}
    \ket{\beta_{\psi}} \to \ket{\zeta}\otimes \underbrace{\Big(\ket{\tilde k_g^x} \ket{\tilde k_g^y}\ket{\tilde c_g}\Big)}_{\ket{g}} \otimes \underbrace{\Big(\ket{\tilde k_q^x} \ket{\tilde k_q^y}\ket{\tilde c_q}\Big)}_{\ket{q}}.
\end{align}
We combine the longitudinal encoding with the gluon occupancy using the quantum register $\ket{\zeta}$. 
The $\zeta=0$ state encodes the $\ket{q}$ state with $k^+ = K$; and $\zeta=[k_g^+]_{\text{bin}}$ encodes the $\ket{qg}$ states with $k_g^+=\{1,2,\ldots,K-1/2\}$ and $k_q^+=K-k_g^+$. We label the transverse modes by $k_l^i=\{0,1,\ldots, 2N_\perp-1\}$ for $i=x,y$, and map each mode to the qubit state $\tilde k^i_l=[k^i_l]_{\text{bin}}$. In a similar way, the color register encodes the $c_l$ mode to the qubit state $\tilde c_l=[c_l-1]_{\text{bin}}$.
In total, we need a number of $n_Q=\log_2(2^3\, \Kceil\, (2 N_\perp)^4)$ qubits. 

    \item Initial state preparation. We assign the initial state as a single quark state with $ k^x=k^y=0$ and $\ket{\tilde c_q}=1/\sqrt{2}(\ket{0}+\ket{1}) $, which can be prepared by acting on the corresponding transverse register with an X gate, and the color qubit with a Hadamard gate.
    \item Time evolution. We first trotterize the evolution operator into small timestep operations, as in Eq.~\eqref{eq:time_evolution_exp}. 
    Then for each timestep, we implement the product formula in Eq.~\eqref{eq:product_formula}. 
    For each part of the Hamiltonian, we find the corresponding Pauli terms, which we then evolve  using {\tt PauliEvolutionGate} class provided by {\tt Qiskit}.
    We apply quantum Fourier Transform in the places of $\mathcal{FT}$ on the circuit. 
    \item Measurement.
    Upon measurement, the jet state on the circuit, as denoted in Eq.~\eqref{eq:basis_expansion}, collapses to a basis state $\ket{\beta}$ with probability $ |v_{\beta} (x^+) |^2$.
    By performing multiple measurements (shots), we are able to reconstruct the distribution of the jet state in the basis space. Quantities including momentum broadening and gluon numbers can be extracted from the distribution.
\end{enumerate}

\section{Results and discussion}\label{sec:res}
\begin{figure}[!t]
\centering
    \subfigure[\label{fig:cross} Cross section of the single quark state]{
    \includegraphics[width=0.48\textwidth]{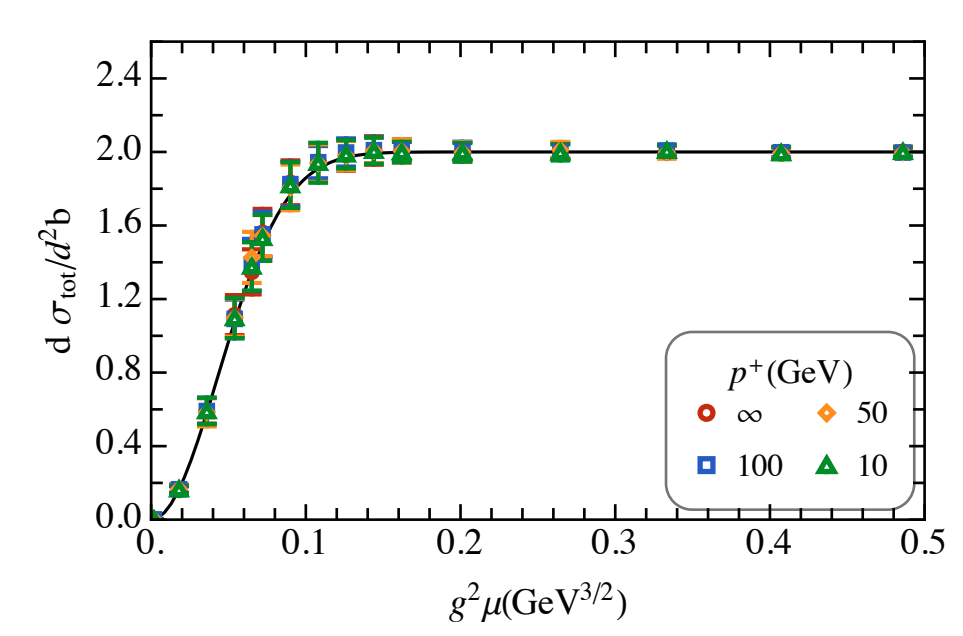}}
    \subfigure[\label{fig:delta_Pqg} Medium modification of gluon emission]{
    \includegraphics[width=0.48\textwidth]{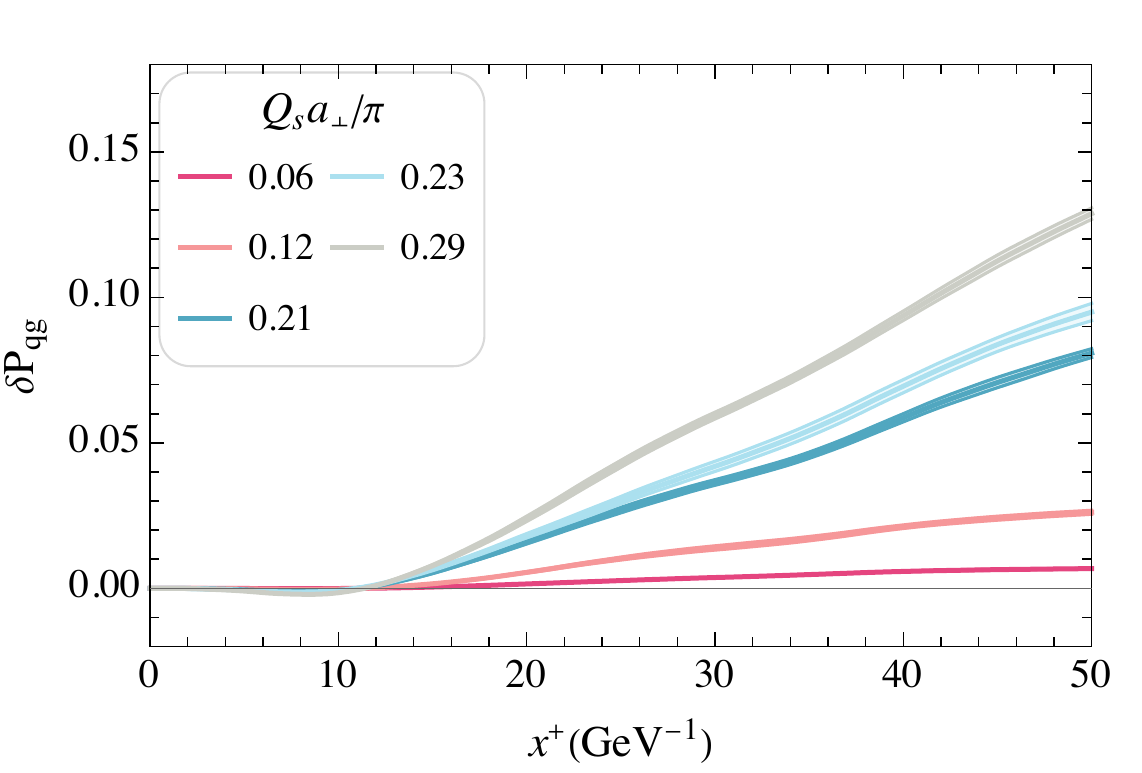}}
    \caption{Results from classical simulation. Figures are adapted from Refs.~\cite{Li:2020uhl, Li:2023jeh}.
    }
    \label{fig:res_classical}
    
\end{figure}
\begin{figure}[htp]
    \centering
        \subfigure[\label{fig:p2_Pplus1} Momentum broadening]{
    \includegraphics[width=0.48\textwidth]{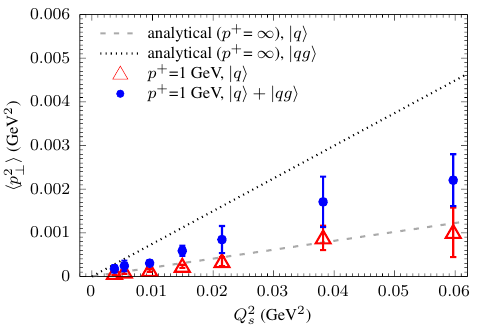}}
        \subfigure[\label{fig:medium_evo} Gluon emission]{
    \includegraphics[width=0.48\textwidth]{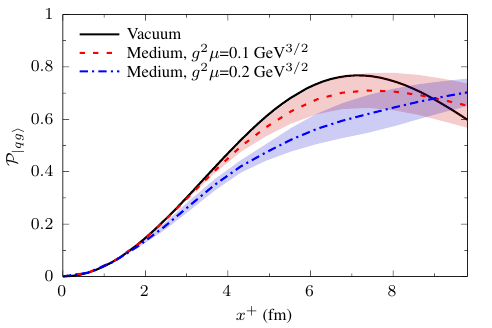}}
    \caption{Results from quantum simulation. Figures are adapted from Ref.~\cite{Barata:2023clv}
    }\label{fig:res_quantum}
\end{figure}
In the classical simulation work using tBLFQ~\cite{Li:2020uhl}, we studied the evolution of a single quark state beyond the eikonal limit by letting it carry a finite $p^+$. Figure ~\ref{fig:cross} shows the total cross section at different medium strength, for various different $p^+$s. 
The agreement between the tBLFQ result at $p^+=\infty$ and the eikonal analytical expectation (shown in the black solid line) helps verify our formalism. Furthermore, the results for finite $p^+$ also agree with that in the eikonal, suggesting that the non-eikonal correction to the cross section is negligible in $\ket{q}$ space.  However, a noticeable non-eikonal effect is observed in the quark's distribution in the transverse position space in the same work. 
In the subsequent works \cite{Li:2021zaw, Li:2023jeh}, we studied jet evolution in the $\ket{q}+\ket{qg}$ Fock space, allowing gluon emission and absorption. 
We ran simulations on the transverse lattice of $N_\perp=16$ and $L_\perp = 50 \GeV^{-1} $, with total longitudinal momentum quanta $K=8.5$, and for a duration of $L_\eta = 50 \GeV^{-1}$. 
In Fig.~\ref{fig:delta_Pqg}, we present the medium induced gluon emission $\delta P_{\ket{qg}}$, the difference of the probability of the quark jet in the $\ket{qg}$ sector in the medium and that in the vacuum.
The different curves at different saturation scale have a similar behavior. It forms a small dip in the early time region, then becomes positive and grows linearly in time.
Comparing the curves at different $Q_s$, one can see that the denser the medium, the more the $\ket{qg}$ component develops.

In the quantum simulation works~\cite{Barata:2022wim, Barata:2023clv}, we studied the jet momentum broadening and the gluon production for various medium strengths. 
We ran the simulation of quark jet in the $\ket{q}$ and $\ket{q}+\ket{qg}$ Fock spaces, on a basis of $N_\perp=1$ and $L_\perp = 32 \GeV^{-1} $, with $K=3.5$ and $L_\eta = 10$ fm. 
The results of the final state $\braket{p^2}$ at various saturation scales $Q_s$ is presented in Fig.~\ref{fig:p2_Pplus1}.
The result of the single quark state is close to the corresponding eikonal expectation, whereas the $\ket{q}+\ket{qg}$ result deviates away from the single quark's result, indicating non-eikonal effects due to gluon emission.
We compared the gluon emission probability, $\mathcal P_{\ket{qg}}$, in vacuum and in the medium in Fig.~\ref{fig:medium_evo}. The presence of the medium suppresses the gluon emission at early time, and enhances it at later times, a behavior similar to that observed in the classical simulation in Fig.~\ref{fig:delta_Pqg}.

\section{Summary and outlook}\label{sec:sum}
Based on the light-front Hamiltonian formalism, we have developed a nonperturbative computational framework to simulate the quark jet evolution, first on classical computers, and subsequently on quantum simulators. 
The main advantage of quantum simulation lies in its logarithmic reduction of the computational resources provided by the qubit encoding scheme.
Extensions of the quantum algorithm to higher Fock sectors with a more efficient strategy are underway, enabling us to study jet as a many-particle quantum state from first principles.

\section*{Acknowledgments}
We are grateful to J. Barata, G. Chen, X. Du, T. Lappi, Y. Li, P. Maris, W. Qian, C. A. Salgado, K. Tuchin, J. P. Vary, X. Zhao, who has made important contributions to this work.
M.L. is supported by Xunta de Galicia (CIGUS accreditation), European Union ERDF, the Spanish Research State Agency under project PID2020-119632GB-I00, and European Research Council under project ERC-2018-ADG-835105 YoctoLHC. 

\bibliographystyle{JHEP}
\bibliography{qA}

\providecommand{\href}[2]{#2}\begingroup\raggedright\begin{thebibliography}{1}

\bibitem{Li:2020uhl}
M.~Li, X.~Zhao, P.~Maris, G.~Chen, Y.~Li, K.~Tuchin et~al., \emph{{Ultrarelativistic quark-nucleus scattering in a light-front Hamiltonian approach}}, \href{https://doi.org/10.1103/PhysRevD.101.076016}{\emph{Phys. Rev. D} {\bfseries 101} (2020) 076016} [\href{https://arxiv.org/abs/2002.09757}{{\ttfamily 2002.09757}}].

\bibitem{Li:2021zaw}
M.~Li, T.~Lappi and X.~Zhao, \emph{{Scattering and gluon emission in a color field: A light-front Hamiltonian approach}}, \href{https://doi.org/10.1103/PhysRevD.104.056014}{\emph{Phys. Rev. D} {\bfseries 104} (2021) 056014} [\href{https://arxiv.org/abs/2107.02225}{{\ttfamily 2107.02225}}].

\bibitem{Li:2023jeh}
M.~Li, T.~Lappi, X.~Zhao and C.A.~Salgado, \emph{{Momentum broadening of an in-medium jet evolution using a light-front Hamiltonian approach}}, \href{https://doi.org/10.1103/PhysRevD.108.036016}{\emph{Phys. Rev. D} {\bfseries 108} (2023) 036016} [\href{https://arxiv.org/abs/2305.12490}{{\ttfamily 2305.12490}}].

\bibitem{Barata:2022wim}
J.a.~Barata, X.~Du, M.~Li, W.~Qian and C.A.~Salgado, \emph{{Medium induced jet broadening in a quantum computer}}, \href{https://doi.org/10.1103/PhysRevD.106.074013}{\emph{Phys. Rev. D} {\bfseries 106} (2022) 074013} [\href{https://arxiv.org/abs/2208.06750}{{\ttfamily 2208.06750}}].

\bibitem{Barata:2023clv}
J.a.~Barata, X.~Du, M.~Li, W.~Qian and C.A.~Salgado, \emph{{Quantum simulation of in-medium QCD jets: Momentum broadening, gluon production, and entropy growth}}, \href{https://doi.org/10.1103/PhysRevD.108.056023}{\emph{Phys. Rev. D} {\bfseries 108} (2023) 056023} [\href{https://arxiv.org/abs/2307.01792}{{\ttfamily 2307.01792}}].

\bibitem{1stBLFQ}
J.P.~Vary, H.~Honkanen, J.~Li, P.~Maris, S.J.~Brodsky, A.~Harindranath et~al., \emph{{Hamiltonian light-front field theory in a basis function approach}}, \href{https://doi.org/10.1103/PhysRevC.81.035205}{\emph{Phys. Rev.} {\bfseries C81} (2010) 035205} [\href{https://arxiv.org/abs/0905.1411}{{\ttfamily 0905.1411}}].

\bibitem{McLerran:1993ni}
L.D.~McLerran and R.~Venugopalan, \emph{{Computing quark and gluon distribution functions for very large nuclei}}, \href{https://doi.org/10.1103/PhysRevD.49.2233}{\emph{Phys. Rev.} {\bfseries D49} (1994) 2233} [\href{https://arxiv.org/abs/hep-ph/9309289}{{\ttfamily hep-ph/9309289}}].

\bibitem{Barata:2021yri}
J.a.~Barata and C.A.~Salgado, \emph{{A quantum strategy to compute the jet quenching parameter $\hat{q}$}}, \href{https://doi.org/10.1140/epjc/s10052-021-09674-9}{\emph{Eur. Phys. J. C} {\bfseries 81} (2021) 862} [\href{https://arxiv.org/abs/2104.04661}{{\ttfamily 2104.04661}}].

\end{thebibliography}\endgroup

\end{document}